\documentclass[conference]{IEEEtran}
\usepackage{array}
\usepackage{amssymb}
\usepackage{amsmath}
\usepackage{graphicx}
\usepackage{float}

\newcommand{\myspace}{\;\;}
\DeclareGraphicsExtensions{.eps}

\begin{document}
%
\title{Transmission Policies for Interference Management in Full-Duplex D2D Communication}

\author{

 \IEEEauthorblockN{
  Nikolaos Giatsoglou\IEEEauthorrefmark{1}, Angelos Antonopoulos\IEEEauthorrefmark{2}, Elli Kartsakli\IEEEauthorrefmark{1}, John Vardakas\IEEEauthorrefmark{1} and Christos Verikoukis\IEEEauthorrefmark{2}
 }
 \IEEEauthorblockA{
  \IEEEauthorrefmark{1}IQUADRAT Informatica S.L., Barcelona, Spain\\
  \IEEEauthorrefmark{2}Telecommunications Technological Centre of Catalonia (CTTC), Castelldefels, Spain\\
  Email: \{ngiatsoglou, ellik, jvardakas\}@iquadrat.com, \{aantonopoulos, cveri\}@cttc.es
 }
}

\maketitle

\begin{abstract}
Full-Duplex (FD) wireless and Device-to-Device (D2D) communication are two promising technologies that aspire to enhance the spectrum and energy efficiency of wireless networks, thus fulfilling key requirements of the 5\textsuperscript{th} generation (5G) of mobile networks. Both technologies, however, generate excessive interference, which, if not managed effectively, threatens to compromise system performance. To this direction, we propose two transmission policies that enhance the communication of two interfering FD-enabled D2D pairs, derived from game theory and optimization theory. The game-theoretic policy allows the pairs to choose their transmission modes independently and the optimal policy to maximize their throughput, achieving significant gains when the pairs interfere strongly with each other. 
\end{abstract}


\IEEEpeerreviewmaketitle

\section{Introduction}\label{sec:intro}

The bandwidth demands of the mobile users are always increasing, since they want to enjoy the latest applications of the smartphone community, share multimedia content and be online on a daily basis. In turn, operators want to satisfy their customers, by evolving their infrastructure towards the hyped 5\textsuperscript{th} generation (5G) of mobile networks, which will enhance the energy efficiency and address the spectrum shortage of current mobile communications \cite{IEEEhowto:5g1}. Two promising technologies that aspire to achieve these goals are wireless \emph{full-duplex} and \emph{device-to-device} communication.

Full-Duplex (FD) allows a user to simultaneously transmit and receive in the same frequency band. Initially, this technology was deemed impractical for wireless transceivers, because of the huge \emph{self-interference} leaking from the user's transmitter directly to his receiver, but recent experimental works \cite{IEEEhowto:phy1, IEEEhowto:phy2} managed to sufficiently attenuate the self-interference signal to realize wireless FD communication. This achievement sparked new interest in FD technology, which promised to double the capacity of wireless networks. Although the throughput of a wireless link can be doubled with FD, if the self-interference is adequately suppressed, the performance of FD in a network is significantly reduced, since the interference imprint on the network is doubled, limiting spatial reuse \cite{IEEEhowto:fdquestion1}.

Device-to-Device (D2D) communication allows users to communicate directly to offload traffic from the base station and reduce the energy consumption of the system \cite{IEEEhowto:d2dsurvey}. \textit{In-band} D2D has received particular attention, since it permits the D2D users to reuse the cellular spectrum, and it is further categorized as \textit{overlay}, when the base stations reserve a portion of the cellular bandwidth for D2D traffic, or as \textit{underlay}, when the D2D users compete with the primary cellular users for resources. However, in-band D2D introduces additional interference, which must be carefully addressed to realize the potentials of D2D technology.

From the above descriptions, it is seen that interference limits the performance of both technologies and must be managed effectively to guarantee performance. In the literature, some initial MAC protocols have been proposed that apply to FD communications. \cite{IEEEhowto:mac1,IEEEhowto:mac2} extend traditional HD protocols such as CSMA to the FD paradigm, but this is clearly sub-optimal. \cite{IEEEhowto:mac3} considers an ad-hoc ALOHA network and analyses its performance with the mathematical tool of stochastic geometry. Since the performance bottleneck of an FD network lies in the selfish behavior of the users, it is better suited to a game-theoretic model. Game theory has been considered in \cite{IEEEhowto:mac4}, using the spatially averaged throughput of an ad-hoc network as utility, but the analysis is applied only to an HD network.

Motivated by these observations, in this paper we propose two interference-aware transmission policies, that enable two interfering D2D pairs to enhance their throughput. The game theoretic policy is derived modeling the interactions of the pairs as a non-cooperative game and yields a distributed MAC protocol, where the pairs choose their transmission modes independently based on a simple threshold test. The optimal policy is achieved through cooperation between the pairs, which increases the implementation complexity but has significant throughput gains when the level of interference is high. Our analysis allows us to highlight the challenges and reflect important points on the application of FD technology.

Our paper is structured as follows. In Section~\ref{sec:system}, we present the system model and describe the physical layer and the network layer aspects that are pertinent to our analysis. In Section~\ref{sec:derivations}, we derive the throughput of the D2D pairs, which will be used  in the analysis of the next section as a performance metric. In Section~\ref{sec:analysis}, we introduce two transmission policies, based on a game-theoretic formulation and an optimization problem, and compare their performances. In Section~\ref{sec:conclusion}, we summarize our results and suggest future lines of work. 

\section{System Model}\label{sec:system}
\begin{figure}[t]
\centering
\includegraphics[scale=1]{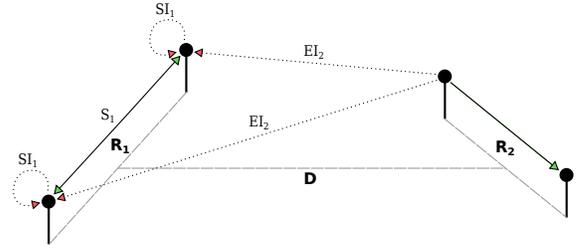}
\caption{System model: a realization where pair $1$ operates in FD mode and pair $2$ in HD mode. Only the interference at pair $1$ is depicted for clarity.}\label{fig:system}
\end{figure}

We consider two FD-enabled D2D pairs, as shown in Fig.~\ref{fig:system}. The two pairs span distances $R_1$ and $R_2$, have arbitrary orientation\footnote{Please note that Fig. \ref{fig:system} depicts a specific network realization.} and are separated by distance $D$, defined as the distance of the midpoints of $R_1$ and $R_2$. They operate in the same frequency band, overlaying primary cellular transmissions, so that the D2D users interfere with each other, but not with the primary cellular users. The users of each pair want to exchange packets and use a slotted protocol to synchronize their transmissions. In each slot, a pair can operate in \textit{Idle}, \textit{Half-Duplex} (\textit{HD}) or \textit{Full-Duplex} (\textit{FD}) mode when it transmit 0, 1 or 2 packets respectively, and we denote the transmission modes of the two pairs as $a_1$ and $a_2$.

To calculate the success probability of a packet transmission and, eventually, throughput, we need to introduce the signal model of our system. We assume that the users of pair 1 and pair 2 transmit with powers $P_1$ and $P_2$ respectively, and that their signals are degraded by power-law path-loss and Rayleigh fading. With these definitions, we express the \emph{Signal-to-Interference-Ratio} ($SIR$)\footnote{Thermal noise is neglected in our study, since interference has the most detrimental impact on communication.} at the receiver of pair $1$ as\footnote{All expressions will be derived from the point of view of pair $1$, but can be applied to pair 2, by exchanging the indices of the variables.} 

\begin{equation}
SIR_1 = \frac{S_1}{EI_2+SI_1}. \label{eq:sinr}
\end{equation}
 
\noindent
In the above equation
\begin{itemize}
\item $S_1$ represents the power of the useful signal, received from the other user of pair $1$. It is equal to
\begin{equation}
S_1 = P_1 h_1 R_1^{-\alpha},
\label{modelS}\end{equation}
where $h_1 \! \sim \! Exp(1)$ is the fading random variable, $P_1$ is the power transmitted from the other user of
 pair $1$ and $\alpha$ is the path-loss coefficient.

\item $EI_2$ represents the power of the \emph{external interference} (\textit{EI}), received from pair $2$.
It depends on the transmission mode of pair $2$ and the cross-distances between the users of pair $1$ and pair $2$, which have been approximated with $D$. This assumption is reasonable for real networks, where $D$ is sufficiently longer than $R_1$ and $R_2$. It yields
\begin{equation}
EI_{2} = \left\{
  \begin{array}{ll}
    0 & \text{if} \myspace a_2= \text{idle} \\
    P_{2} h' D^{-\alpha} & \text{if} \myspace a_2= \text{HD} \\
    P_{2} h' D^{-\alpha}+ P_{2} h'' D^{-\alpha} & \text{if} \myspace a_2= \text{FD} 
  \end{array}
\right.,
\label{eq:modelEI}\end{equation}
where $h',h''\! \sim \! Exp(1)$ are the fading random variables of the interference paths, assumed independent to each other and to $h$.
 
\item $SI_1$ represents the power of the \emph{self-interference} (\textit{SI}), received by the user's own transmission when he operates in FD mode. It actually refers to the \textit{remaining} SI, after the receiver cancels a portion of the loop-back signal. It is equal to
\begin{equation}
SI_{1} = \kappa P_1,\quad \kappa \sim Exp(\beta), 
\label{eq:modelSI}\end{equation}
where $\kappa$ is the SI cancellation coefficient, modeled as an exponential random variable with  $\mathbb{E}[\kappa]=1/\beta$. The coefficient $\kappa$ is assumed random, because it incorporates fading, caused by reflections of the transmitted signal back to the user. Since lower values of $\kappa$ imply higher SI attenuation, the reciprocal parameter $\beta=1/\mathbb{E}[\kappa]$ represents the \textit{mean SI attenuation}, which depends on the implementation of the SI cancellation scheme.
\end{itemize} 

Based on this signal model, a transmission is considered successful when the $SIR$ at the receiver stays above some threshold $\theta$ during the whole transmission. The threshold $\theta$ depends on the modulation and coding scheme of the physical layer, and it is related to the minimum \emph{Signal-to-Noise-Ratio} ($SNR$) needed for perfectly reliable communication at rate $\mathcal{R}$. It is given by the Shannon capacity formula
\begin{equation}
\mathcal{R} = \text{log}(1+\theta) \Leftrightarrow  \theta = 2^\mathcal{R}-1,
\label{eq:shannonCapacity}\end{equation}
where $\mathcal{R}$ is normalized over the transmission bandwidth. The capacity formula applies to the \emph{Additive White Gaussian Noise} (\emph{AWGN}) channel, but interference is commonly approximated as Gaussian, which allows substituting $SNR$ with $SIR$.  
 
\section{Throughput Analysis}\label{sec:derivations}
Based on the definitions of Section~\ref{sec:system}, in Section~\ref{subsec:derivations_psucc} we derive the success probabilities of transmission and in Section~\ref{subsec:derivations_throughput} the throughput of the two pairs.
 
\subsection{Derivation of the Success Probabilities}\label{subsec:derivations_psucc}

The success of a packet transmission, $\mathbf{P}_s$, depends on the transmission modes of both pairs, $a_1$ and $a_2$. For the receiver of pair 1, it holds that

\begin{equation}\begin{split}
\mathbf{P}_{s_1} &= \mathbf{P}(SIR_1> \theta)=\mathbb{E}\left[\mathbf{P}(h>\frac{\theta R_1^a (EI_2+SI_1)}{P_1})\right]= \\ 
& \!\!\!\!= \mathbb{E}\left[e^{- \frac{\theta R_1^a (EI_2+SI_1)}{P_1})}\right]=
M_{EI_2}(- \frac{\theta R_1^a}{P_1}) M_{SI_1}(-\frac{\theta R_1^a}{P_1}),
\label{eq:psucc_raw}\end{split}\end{equation}
where 
\begin{equation}
M_{EI_2}(t)=\mathbb{E}\left[e^{t \, EI_2}\right],
\label{eq:mgfEIdef}\end{equation}
\begin{equation}
M_{SI_1}(t)=\mathbb{E}\left[e^{t \, SI_1}\right]
\label{eq:mgfSIdef}\end{equation}
denote the moment generating functions of random variables $EI_2$ and $SI_1$.

\begin{itemize}
\item To calculate \eqref{eq:mgfEIdef}, we need the moment generating function of the exponential fading variable
\begin{equation}
M_h(t)=\mathbb{E}\left[e^{t \, h}\right]=\frac{1}{1-t},\quad t<1.
\label{eq:mgfEXP}\end{equation}
To simplify notation, we also define the helper variable
\begin{equation}
\tau=\theta \frac{P_2}{P_1} (\frac{R_1}{D})^\alpha.
\label{eq:tau}\end{equation}
Applying \eqref{eq:modelEI},\eqref{eq:mgfEXP} and \eqref{eq:tau} to \eqref{eq:mgfEIdef}, we find

\begin{equation}
\begin{split}
M_{EI_2} \left( -\frac{\theta R_1^a}{P_1} \right)= & \left\{
\begin{array}{ll}
1 & \text{if} \myspace a_2= \text{Idle} \\
\mathbb{E}\left[ e^{-h'\tau}\right] & \text{if}\myspace a_2=\text{HD} \\
\mathbb{E}\left[ e^{-h'\tau}\right]\mathbb{E}\left[ e^{-h''\tau}\right]& \text{if}\myspace a_2=\text{FD} \\    
\end{array}
\right. \\
= & \left\{
 \begin{array}{ll}
  1 & \text{if} \myspace a_2= \text{Idle} \\
  \frac{1}{1+\tau} & \text{if} \myspace a_2= \text{HD} \\
  (\frac{1}{1+\tau})^2 & \text{if} \myspace a_2= \text{FD} \\    
 \end{array}
\right.\\
\end{split}\label{eq:mgfEI}
\end{equation}

\item To calculate \eqref{eq:mgfSIdef}, we need to consider the SI model of \eqref{eq:modelSI} only when pair $1$ operates in FD mode, since otherwise $SI_1=0$. We then find

\begin{equation}
\begin{split}
M_{SI_1}(-\frac{\theta R_1^a}{P_1})= & \left\{
\begin{array}{ll}
1 & \text{if} \myspace a_1 \in \text{ \{Idle, HD\}} \\
\mathbb{E}\left[ e^{-\kappa \theta R_1^a }\right] & \text{if} \myspace a_1= \text{FD}
\end{array}\right. \\
= & \left\{
\begin{array}{ll}
1 & \text{if} \myspace a_1 \in \text{ \{Idle, HD\}} \\
\frac{1}{1+\theta R_1^a / \beta} & \text{if} \myspace a_1= \text{FD}
\end{array}
\right.
\end{split}
\label{eq:mgfSI}\end{equation}
\end{itemize}

We can express the success probability of  \eqref{eq:psucc_raw} compactly, introducing the parameters
\begin{gather}
\lambda_1 \triangleq \frac{1}{1+\theta \frac{ R_1^a}{\beta}} \nonumber \\
\mu_1 \triangleq \frac{1}{1+\tau}=\frac{1}{1+\theta \frac{P_2}{P_1} (\frac{R_1}{D})^\alpha}
\label{eq:lm}\end{gather}
and denoting the number of pair 2's transmissions by $n_2$. Combining \eqref{eq:mgfEI}, \eqref{eq:mgfSI} and \eqref{eq:lm} into \eqref{eq:psucc_raw} yields
\begin{equation}
\mathbf{P}_{s_1}=  \left\{ 
 \begin{array}{ll}
    0 & \text{if} \myspace a_1= \text{Idle} \\
    \mu_1^{n_2} & \text{if} \myspace a_1= \text{HD} \\
    \lambda_1 \mu_1^{n_2} & \text{if} \myspace a_1= \text{FD} \\    
 \end{array} \right. .
\label{eq:psucc_final}\end{equation}

Parameters $\lambda$ and $\mu$ are useful because they abstract the physical layer aspects of our system and distinguish the impact of EI and SI on transmission. Let us describe them in more detail at this point. 
\begin{itemize}
\item $\lambda$ encapsulates SI. It depends on the intra-pair distance, the SI attenuation factor, the $SIR$ threshold and the path-loss, but not on the transmitted power. It ranges between 0, when SI is strong, and 1, when SI is fully canceled.
\item $\mu$ encapsulates EI. It coincides with the success probability of an HD transmission when one external user interferes, and becomes $\mu^2$ when two external users interfere. In this sense, an FD transmission doubles the interference imprint on the rest of the network. $\mu$ tends to 0 when the pairs strongly interfere and to 1 when they do not interact.
\end{itemize}

\subsection{Derivation of Throughput}\label{subsec:derivations_throughput}
We define throughput as the number of successful transmissions that a pair accomplishes in one slot and denote it by $\rho$. For player 1 
\begin{itemize}
\item if $a_1=\text{Idle}$
\begin{equation}
\rho_1=0.
\label{eq:rho_noEI}\end{equation}

\item if $a_1=\text{HD}$
\begin{equation}
\rho_1=1 \cdot \mathbf{P}_{s_1}+0 \cdot (1-\mathbf{P}_{s_1})=\mathbf{P}_{s_1}.\label{eq:rho_hdEI}
\end{equation}

\item if $a_1=\text{FD}$
\begin{equation}
\rho_1=\mathbf{P}_{s_1}^{(1)}+\mathbf{P}_{s_1}^{(2)}=2 \mathbf{P}_{s_1},
\end{equation} 
where $\mathbf{P}_{s_1}^{(1)}$ and $\mathbf{P}_{s_1}^{(2)}$ distinguish the success probabilities at the two receivers of pair 1. They are considered equal, based on our assumption that both receivers experience the same mean EI from pair 2.
\end{itemize}
To finish the derivation, we note that $\mathbf{P}_{s_1}$ must be substituted from \eqref{eq:psucc_final}, considering the transmission modes of both pairs.

\begin{table}[b]
\centering
\caption{Throughput game for two interfering D2D pairs}
\newcolumntype{C}{>{\centering\arraybackslash}m{1.8cm}}
\renewcommand{\arraystretch}{1.3}
\begin{tabular}{m{0.2cm} r|C|C|C|}
& \multicolumn{1}{c}{} & \multicolumn{1}{c}{} & \multicolumn{1}{c}{$ \mathbf{P2} $} & \multicolumn{1}{c}{} \\
& \multicolumn{1}{c}{} & \multicolumn{1}{c}{Idle} & \multicolumn{1}{c}{HD} & \multicolumn{1}{c}{FD} \\ \cline{3-5}	
&  Idle & $0,0$ & $0,1$ & $0,2\lambda_2$ \\ \cline{3-5} 
$ \mathbf{P1} $ & HD & $1,0$ & $\mu_1,\mu_2$ & $\mu_1^2, 2\lambda_2\mu_2$ \\\cline{3-5} 
&  FD & $2 \lambda_1,0$ & $2 \lambda_1 \mu_1 ,\mu_2^2 $ & $2 \lambda_1 \mu_1^2, 2 \lambda_2 \mu_2^2$ \\ \cline{3-5}
\end{tabular}
\label{table:game}
\end{table}

\section{Transmission Policies in FD D2D}\label{sec:analysis}
Since the throughput of the D2D pairs depends on interference, in this section we propose two interference-aware transmission policies that shape the interference pattern of the network and enhance performance. In Section~\ref{subsec:analysis_game}, we introduce a game-theoretic policy which forces the network to operate in a stable equilibrium, and in Section~\ref{subsec:analysis_game}, we derive an optimal policy which maximizes throughput. In Section~\ref{subsec:analysis_comparison}, we compare the performance of the two policies.

\subsection{Game-Theoretic Transmission Policy}\label{subsec:analysis_game}
The D2D pairs have an incentive to operate in FD mode to increase their transmission rate, but if all pairs adopt this policy, the excessive interference degrades throughput. This outcome is due to the selfish behavior of the D2D users, who want to transmit without considering the burden that they cause to their neighbors. We model this conflicting situation as a \emph{non-cooperative game}, defined as a tuple $\mathcal{G}=(\mathcal{N},\mathcal{S}_i,\mathcal{U}_i), \myspace i \in \mathcal{N}$, where $\mathcal{N}$ is the set of the players, $\mathcal{S}_i$ is the set of the strategies of player $i$ and $\mathcal{U}_i$ is the utility of player $i$. In our formulation
\begin{gather}
\mathcal{G}=\left( \left\{1,2\right\},\left\{\text{Idle, HD, FD}\right\},\rho_i \right), \quad i \in \{1,2\},
\end{gather}
which is represented in the matrix form shown in Table \ref{table:game}.

\begin{figure}[t]
\centering
\includegraphics[scale=0.5]{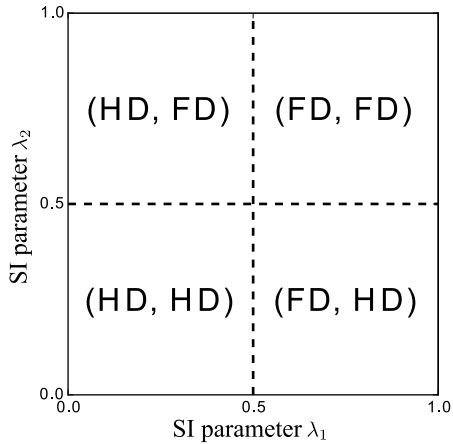}
\caption{Strategy regions for the game of Table \ref{table:game}.}\label{fig:regions_for_NE}
\end{figure}

This game admits a single dominant solution. The reason is that for every player the Idle strategy is always strictly dominated, since it contributes zero to his utility, and between HD and FD, there is a strictly dominant strategy, depending on the value of his SI parameter. For player 1, FD dominates HD when
\begin{subequations}
\begin{equation}
2 \lambda_1 > 1 \myspace \overset{\eqref{eq:lm}}{\Longrightarrow} \myspace \beta > {\theta R_1^\alpha}
\end{equation}
and HD dominates FD when
\begin{equation}
2 \lambda_1 < 1 \myspace \overset{\eqref{eq:lm}}{\Longrightarrow} \myspace  \beta < {\theta R_1^\alpha}.
\end{equation}
\label{eq:criterion}\end{subequations}
Since $\beta$ represents the mean SI attenuation, \eqref{eq:criterion} instructs player 1 to play FD when his SI cancellation is adequate and HD otherwise. This criterion also applies to player 2, so it determines one strictly dominant solution, which constitutes the unique pure \textit{Nash Equilibrium (NE)} of the game. The uniqueness of the dominant solution precludes the existence of a NE in mixed strategies. Fig. \ref{fig:regions_for_NE} visualizes the four possible combinations of strategies for the NE.

Since the players choose their strategies without considering the parameters of their opponents, the NE yields a distributed transmission policy, which causes the pairs to transmit in every time slot. We denote these pure strategies

\begin{table}[H]
\normalsize
\centering
\begin{tabular}{lcl}
\textit{pure HD} & : & pair operates continuously in HD mode\\
\textit{pure FD }& : & pair operates continuously in FD mode.\\
\end{tabular}
\end{table}

\begin{figure}[t]
\centering
\includegraphics[scale=0.5]{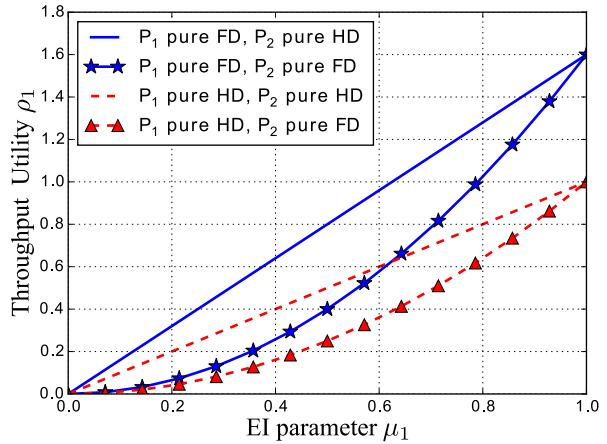}
\caption{Throughput of player $1$ when $\lambda_1=0.8$ for every non-idle strategy combination of player 1 and 2.}\label{fig:game}
\end{figure}

We now illustrate the effect of the game-theoretic transmission policy on the utility of player 1. Fig. \ref{fig:game} depicts the throughput of player 1 as a function of the EI parameter $\mu_1$ for every candidate NE. We assume $\lambda_1=0.8>0.5$ to focus on the FD technology. We observe that the performance drops when player 2 switches from pure~HD to pure~FD, which is justified by the doubling of EI. We also verify that, regardless of the strategy of player 2, player 1 responds with pure~FD which leads to better throughput, as predicted by the NE. Another important observation is that the strategy combination (pure~HD,~pure~HD) outperforms (pure~FD,~pure FD) when $\mu_1$ is low, even though the SI parameter $\lambda_1$ favors pure FD. The reason is that the good SI cancellation is overshadowed by the strong EI that the pairs generate, possibly due to proximity or high transmitter power. In this case, the optimal strategy combination (pure~HD,~pure~HD) cannot be achieved when players operate independently and some form of cooperation is needed instead.

\begin{table}[b]
\newcolumntype{C}{>{\centering\arraybackslash}m{1.8cm}}
\centering
\caption{Pair throughput for a symmetric configuration}
\renewcommand{\arraystretch}{1.3}
\begin{tabular}{m{0.2cm} r|C|C|C|}
& \multicolumn{1}{c}{} & \multicolumn{1}{c}{} & \multicolumn{1}{c}{$ \mathbf{P2} $} & \multicolumn{1}{c}{} \\
& \multicolumn{1}{c}{} & \multicolumn{1}{c}{Idle} & \multicolumn{1}{c}{HD} & \multicolumn{1}{c}{FD} \\ \cline{3-5}		
&  Idle & $0,0$ & $0,1$ & $0,2\lambda$ \\ \cline{3-5} 
$ \mathbf{P1} $ & HD & $1,0$ & $\mu,\mu$ & $\mu^2, 2\lambda\mu$ \\\cline{3-5} 
&  FD & $2 \lambda,0$ & $2 \lambda \mu ,\mu^2 $ & $2 \lambda \mu^2, 2 \lambda \mu^2$ \\ \cline{3-5}
\end{tabular}
\label{table:opt}
\end{table}
 
\subsection{Optimal Transmission Policy}\label{subsec:analysis_optimization}
In Section \ref{subsec:analysis_game}, we observed that the NE is not necessarily optimal when EI is strong and $\mu_1$ and $\mu_2$ are close to 0. In this case, the pairs can gain by mutually agreeing to mix their strategies with Idle to sporadically alleviate EI. We introduce a new transmission policy which permits the two pairs to operate in Idle, HD and FD mode with probabilities $p_0$, $p_1$ and $p_2$ respectively and optimize its performance. We consider the symmetric case, where 
\begin{gather}
\lambda_1=\lambda_2=\lambda\nonumber\\
\mu_1=\mu_2=\mu.
\end{gather}
and summarize the throughput in Table \ref{table:opt}, modifying Table \ref{table:game} of Section \ref{subsec:analysis_game}. We then formulate the optimization problem 
\begin{gather}
\textbf{maximize } \rho = \sum_{i,j} p_i p_j \mathcal{U} (a_i ,a_j) = \nonumber\\
 p_0 p_1 + \mu p_1^2 +\mu^2 p_1 p_2 + 2\lambda p_2 p_0 + 2\lambda \mu p_2 p_1 +2\lambda \mu^2 p_2^2 \nonumber\\
 \textbf{s.t. } p_0+p_1+p_2=1  \nonumber\\
  0<p_0,p_1,p_2<1
\label{eq:optproblem}\end{gather}

To solve \eqref{eq:optproblem}, we substitute $p_2 = 1-p_1-p_0$ and locate the points $(p_0,p_1)$, where the gradient is zero
\begin{equation}
\nabla \rho = (\frac{\partial \rho}{\partial p_0}, \frac{\partial \rho}{\partial p_1})= (0,0).
\label{opt}\end{equation}
Solving \eqref{opt} yields no local maximum inside the domain of $(p_0,p_1)$, but we must still check the three boundaries, which correspond to the cases $p_0=0$, $p_1=0$ and $p_2=0$ respectively. To distinguish these mixed strategies from the pure strategies of Section \ref{subsec:analysis_game}, we denote them

\begin{figure}[t]
\centering
\includegraphics[scale=0.35]{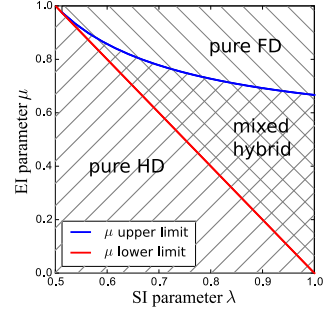}
\caption{The limits of $\mu$, which demarcate the mixed hybrid region.}\label{fig:regions_for_hybrid}
\end{figure}

\begin{table}[H]
\normalsize
\centering
\begin{tabular}{lcl}
\textit{mixed HD} & : & pair mixes HD and Idle mode \\
\textit{mixed FD }& : & pair mixes FD and Idle mode\\
\textit{mixed hybrid} & : & pair mixes HD and FD mode
\end{tabular}
\end{table}
\noindent 
Continuing with the analysis
\begin{itemize}
\item if $p_2=0$ (mixed HD)
\begin{equation}
\rho= p_0 p_1 + \mu p_1^2=p_1 (1-(1-\mu)p_1).
\end{equation}

This is maximized at 
\begin{equation}
p_1= \left\{
  \begin{array}{ll}
    \frac{1}{2(1-\mu)} & : \mu < 1/2 \\
    1 &: \mu > 1/2 
  \end{array}
\right.,
\end{equation}

yielding  
\begin{equation}
\rho_{max} = \left\{
  \begin{array}{ll}
    \frac{1}{4(1-\mu)} & : \mu < 1/2 \\
    \mu &: \mu > 1/2 
  \end{array}
\right..
\label{eq:rho_max_mixed_HD}\end{equation}

\item if $p_1=0$ (mixed FD)
\begin{equation}
\rho = 2\lambda p_2 p_0 + 2\lambda \mu^2 p_2^2=2\lambda p_2 (1-(1-\mu^2)p_2).
\end{equation}

This is maximized at 
\begin{equation}
p_2= \left\{
  \begin{array}{ll}
    \frac{1}{2(1-\mu^2)} & : \mu < 1/\sqrt{2} \\
    1 &: \mu > 1/\sqrt{2}
  \end{array}
\right.,
\end{equation}

yielding
\begin{equation}
\rho_{max} = \left\{
  \begin{array}{ll}
    \frac{\lambda}{2(1-\mu^2)} & : \mu < 1/\sqrt{2} \\
    2 \lambda \mu^2 &: \mu > 1/\sqrt{2}
  \end{array}
\right..
\label{eq:rho_max_mixed_FD}\end{equation}

\item if $p_0=0$ (mixed hybrid)
\begin{equation}
\rho =  \mu p_1^2 +\mu^2 p_1 p_2 + 2\lambda \mu p_2 p_1 +2\lambda \mu^2 p_2^2.
\label{eq:rho_mixed_hybrid}\end{equation}

This is maximized at
\begin{equation}
p_1= \left\{
  \begin{array}{ll}
	1 & \mu<2(1-\lambda)\\
    \frac{4\lambda \mu-2\lambda-\mu}{2(1-2\lambda)(1-\mu)} & 2(1-\lambda)<\mu<\frac{2\lambda}{4\lambda-1} \\
    0 & \mu>\frac{2\lambda}{4\lambda-1} \\
  \end{array}
\right.,
\label{eq:p1_mixed_hybrid}\end{equation}
where the limits of $\mu$ are valid only when 
\begin{equation}
0<2(1-\lambda)<\frac{2\lambda}{4\lambda-1}<1.
\label{eq:condition_mixed_hybrid}\end{equation}
Fig. \ref{fig:regions_for_hybrid} shows the lower and upper limit of $\mu$ as a function of $\lambda$. For $\lambda<0.5$ both limits exceed 1, so \eqref{eq:condition_mixed_hybrid} does not hold and the mixed hybrid strategy simplifies to pure HD from the game-theoretic policy. For $\lambda>0.5$, we can calculate the maximum throughput by plugging \eqref{eq:p1_mixed_hybrid} in \eqref{eq:rho_mixed_hybrid}, but it is not presented due to space limitations. 
\end{itemize} 

The three boundary maxima must be compared to find the global maximum, which is straight-forward but analytically not insightful, because of the branching of the throughput expressions. We illustrate instead the throughput of each mixed strategy with diagrams. We study separately the cases $\lambda>0.5$ and $\lambda<0.5$, which were highlighted in the game-theoretic policy, to facilitate the comparison of the two policies in the next section.

In Fig. \ref{fig:superlambda}, we plot the throughput for $\lambda=1$ and $\lambda=0.6$. For $\lambda=1$, mixed FD outperforms all other strategies, while for $\lambda=0.6$, the optimal strategy depends on the EI parameter $\mu$. Specifically, mixed FD performs better than mixed HD when $\mu$ is close to 0 and close to 1, but mixed HD is superior in the intermediate region. Mixed hybrid does not contribute to the optimal performance and its throughput approaches zero when EI is severe in both cases. This fact corroborates that the pairs should restrain their transmissions when they interfere strongly, rather than transmit continuously.

\begin{figure}[t]
\centering
\includegraphics[scale=0.35]{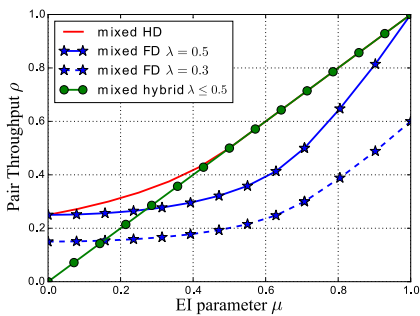}
\caption{Optimal Pair Throughput for $\lambda \leq 0.5$. Two cases are shown: $\lambda=0.5$ and $\lambda=0.3$. The mixed hybrid case is the same $\forall \; \lambda \leq 0.5$. }\label{fig:sublambda}

\end{figure}

In Fig. \ref{fig:sublambda}, we plot the throughput for $\lambda=0.5$ and $\lambda=0.3$. The choice of $\lambda$ actually affects only the performance of mixed FD, since mixed HD does not incur SI, and mixed hybrid coincides with pure HD (no Idle mode) for $\lambda \leq 0.5$, as mentioned in the analysis. We observe that mixed HD outperforms all other strategies, regardless of $\lambda$ and $\mu$, since SI takes its toll on transmission. At $\mu$ close to 0, the throughput of mixed hybrid approaches zero, as in the $\lambda>0.5$ case, and is outperformed by mixed FD, which again highlights the importance of mixing transmissions with Idle when interference is severe.

\subsection{Comparison}\label{subsec:analysis_comparison}
Before we compare the performance of the two transmission policies, let us summarize their characteristics. The game-theoretic policy of Section \ref{subsec:analysis_game} yielded two pure transmission strategies (i.e., pure HD and pure FD), which instruct the pairs to transmit continuously in HD or FD mode respectively. The optimal policy of Section \ref{subsec:analysis_optimization} yielded three mixed transmission strategies (i.e., mixed HD, mixed FD and mixed hybrid), which instruct the players to mix two transmission modes with an optimal probability distribution. The game-theoretic policy is distributed and instructs each player to independently choose his strategy, performing a threshold test on his SI parameter $\lambda$, while the optimal policy requires cooperation and encourages the users to mutually choose their strategies, according to the common experienced EI parameter $\mu$. In the rest of the section we will not consider the mixed hybrid strategy of the optimal policy, as it had inferior performance compared to the other mixed strategies in all considered cases.

\begin{figure}[t]
\centering
\includegraphics[scale=0.35]{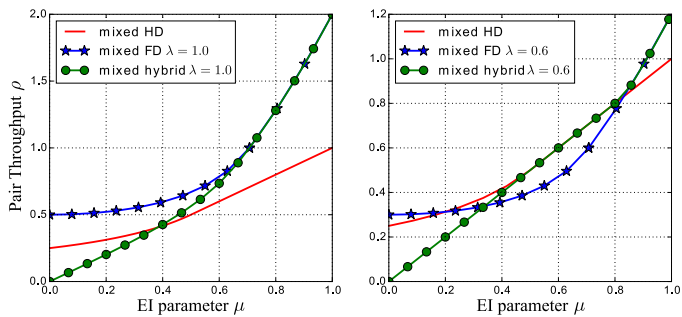}
\caption{Optimal Pair Throughput for $\lambda>0.5$. Two cases are shown: $\lambda=1$ (perfect SI cancellation) and $\lambda=0.6$.}\label{fig:superlambda}
\end{figure}

\begin{figure}[t]
\centering
\includegraphics[scale=0.35]{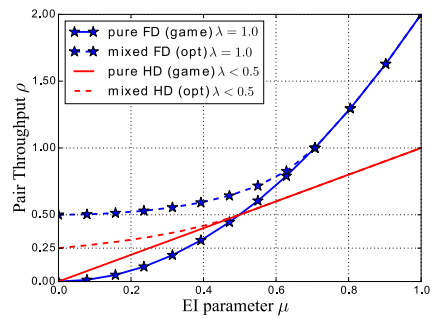}
\caption{Comparison of the game-theoretic and the optimal transmission policy for $\lambda=1$ and $\lambda<0.5$}\label{fig:comparison}
\end{figure}

To compare the two policies, Fig. \ref{fig:comparison} depicts their throughput for a case with low SI ($\lambda>0.5$) and a case with high SI ($\lambda<0.5$). For the low SI case, we choose $\lambda=1$, which implies perfect SI cancellation and favors FD regardless of EI, as shown in Fig. \ref{fig:superlambda}. For the high SI case, the choice of $\lambda$ is irrelevant, since the resulting strategy is always HD and its throughput is not affected by SI. These choices allow us to compare the FD to the HD mode, as well as the two transmission policies.

In  Fig. \ref{fig:comparison}, we observe that the optimal mixed strategies coincide with the pure strategies for $\mu>1/2$ in the HD case and $\mu>1/\sqrt{2}$ in the FD case, as shown in the throughput expressions \eqref{eq:rho_max_mixed_HD} and \eqref{eq:rho_max_mixed_FD}. For smaller $\mu$, the gains of optimization become apparent, as the throughput of the game policy is driven to zero, while the optimal policy still permits some precious throughput. For $\mu=0$ the gain of mixed FD is 0.5 and the gain of mixed HD is 0.25. We also observe that, when EI is strong, pure FD has inferior performance than both the pure HD and the mixed HD strategy, despite perfect cancellation of SI. FD is unconditionally superior to HD, only in the optimal mixed FD mode, which indicates that FD technology should be used carefully to realize its promised performance advantages.

\section{Conclusion}\label{sec:conclusion}
We considered the communication of two FD-enabled D2D pairs, whose interference limits their throughput. We proposed two interference-aware transmission policies, derived by formulating their interactions as a non-cooperative game and as an optimization problem. The game-theoretic policy yields a distributed MAC protocol, which enables the pairs to independently choose their transmission modes based on a simple threshold test. The optimal strategy requires cooperation from the D2D pairs, but offers significant gains in throughput when the interference between the pairs is high. In future work, we plan to extend our game formulation to the N-pair case.

\section*{Acknowledgements}
This work has been funded by the Research Projects AGAUR DI-2015, AGAUR 2014-SGR-1551, CellFive (TEC2014-60130-P) and 5Gwireless (641985).

\end{document}